# Nanoscale crack propagation in clay with water adsorption through reactive MD modeling

Zhe Zhang, Xiaoyu Song*

*Engineering School of Sustainable Infrastructure and Environment, University of Florida, Gainesville, Florida*

**Abstract**

The atomic-scale cracking mechanism in clay is vital in discovering the cracking mechanism of clay at the continuum scale in that clay is a nanomaterial. In this article, we investigate mechanisms of mode I and mode II crack propagations in pyrophyllite and Ca-montmorillonite with water adsorption through reactive molecular dynamics with a bond-order force field. Clay water adsorption is considered by adding water molecules to the clay surface. During the equilibration stage, water adsorption could cause bending deformation of the pre-defined edge crack region. The relatively small orientating angle of water molecules indicates the formation of hydrogen bonds in the crack propagation process. The peak number density of adsorbed water decreases with the increasing strains. The atomistic structure evolution of the crack tip under loading is analyzed to interpret the nanoscale crack propagation mechanism. The numerical results show that the crack tip first gets blunted with a significant increase in the radius of the curvature of the crack tip and a slight change in crack length. The crack tip blunting is studied by tracking the crack tip opening distance and O-Si-O angle in the tetrahedral Si-O cell in mode I and II cracks. We compare bond-breaking behaviors between Al-O and Si-O. It is found that Si-O bond breaking is primarily responsible for crack propagation. The critical stress intensity factor and critical energy release rate are determined from MD simulation results.

*Keywords:* Clay, Crack propagation, Water adsorption, Molecular dynamics, Reactive force field

## 1. Introduction

Cracking in clay could significantly compromise the integrity of the dam and embankment and trigger failure of soil slopes and foundations on clay (e.g., [1–7]). Because clay is a nanomaterial by nature [8], the nanoscale cracking mechanism in clay particles plays a significant role in discovering the cracking mechanism of clay minerals at the continuum and field scales. Water can be strongly attracted to clay mineral surfaces and affect the mechanical and physical properties of clay at multiple scales [3]. The clay-water adsorption can be originated from two primary mechanisms, short-ranged physicochemical mechanisms and capillary condensation at higher relative humidity values [9]. At the nanoscale, physicochemical mechanisms include surface hydration on clay mineral surfaces at low relative humidity, exchangeable cation hydration, and interlayer swelling for expandable clay minerals. The inter-atomic and inter-particle forces could affect physical and chemical bonds such as covalent, ionic, and van der Waals bonds in soil with water [10]. Different structures and properties of adsorbed water are associated with different mechanisms of clay-water interaction [11]. The density and thickness of the adsorbed water layer could be a function of water content and the type of clay [12]. The arrangement/orientation of adsorbed water differs significantly from the bulk water. For example, the formation of strong hydrogen bonds could alter the layout of water molecules [13, 14]. Studies have also clarified the structure of electrical double layers formed on hydrated clay mineral surfaces [15]. Sposito et al. [16] investigated the effect of cation size and charge on the coordination of interlayer cations, interlayer water molecules, and clay mineral surface oxygens. The increasing interlayer cation can enhance the influence of clay mineral structure and hydrophobicity on the configurations of adsorbed water. From the above brief literature review, it may be arguably agreed that the clay water adsorption mechanism has been well studied [10, 17]. However, the crack propagation mechanism in clay under water adsorption remains an open question. To help close this knowledge gap, in this study, we hypothesize that water adsorption on clay surfaces could affect the crack propagation mechanism in clay particles under mode I and mode II cracking at the nanoscale. To test this hypothesis, we conduct reactive molecular dynamics simulations of two clay minerals (pyrophyllite and Ca-montmorillonite), incorporating water adsorption with a bond order reactive force field for clay. Next, we briefly review the computational modeling of clay, including clay cracking, at the nanoscale through molecular dynamics.

---
*Corresponding author
*Email address:* xysong@ufl.edu (Xiaoyu Song)

With rapid advances in high-performance computing, molecular dynamics modeling has become a viable tool to probe the nanoscale mechanics and physics of materials, including clays (e.g., [18–21]). Indeed, numerous studies on modeling the mechanics and physics of clays using molecular dynamics have been reported in the literature (e.g., [22–26], among many others). For instance, Sato et al. [27] and Militzer et al. [28] conducted first-principles studies on the elastic constants of clay minerals based on density functional theory. Zartman et al. [29] studied the tensile moduli, shear moduli, and failure mechanisms for pyrophyllite, montmorillonite, and mica under a wide range of stress conditions through MD simulations. Teich-McGoldrick et al. [30] performed MD simulations of structural and mechanical properties of muscovite under the effect of pressure and temperature. Hantal et al. [31] studied the failure properties of illite with pre-existing cracks within clay layers through MD. However, in the studies mentioned above, dry clay was usually employed without considering water adsorption. A few researchers have investigated the mechanical properties of unsaturated clay with varying hydration by MD simulation [32, 33]. Ebrahimi et al. [34] investigated the anisotropic elastic properties of Na-montmorillonite with water absorption through MD simulations. Carrier et al. [35] calculated elastic properties of swelling montmorillonite as a function of water content. Qomi et al. [36] calculated the nanoscale cohesion, friction angle, cohesion, and hardness of Na-montmorillonite clay with one water layer. Jia et al. [37] simulated hydraulic fracturing in kaolinite at the nanoscale through molecular dynamics. Zhe and Song [38] recently investigated the initiation and growth of cracks in dry clay sheets through molecular dynamics with a general clay force field. However, to the best of our knowledge, no study has been conducted to investigate the mechanism of crack propagation in clay platelets incorporating water adsorption at the atomic scale. Thus, in this article, we conduct MD modeling of crack propagation in clay platelets with water adsorption through reactive MD modeling. In what follows, we review the bond-order reactive force fields for modeling bond breakage during the crack propagation in clay particles using reactive molecular dynamics with clay-water interaction. We refer to the literature (e.g., [21, 22, 39, 40]) for a review of general force fields for modeling the mechanics and physics of clay using MD.

Bond-order force field is a class of interatomic potentials used in reactive MD simulations, e.g., Tersoff [41], REBO [42], AIREBO [43], and ReaxFF [44]. ReaxFF has been utilized to investigate fracture properties of materials such as silicon [45] and graphyne [46]. Zhang et al. [47] examined the stress corrosion process of strained quartz in liquid water and showed the primary role of Si-O bond breakage in crack growth. Buehler et al. [45] investigated the crack limiting speed, crack instabilities, and directional dependence on crystal orientation in a single silicon crystal through MD modeling with ReaxFF. Hou et al. [48] interpreted the disconnecting role of water by studying the tensile failure mechanism of C-S-H gels at different humidity levels. Rimsza et al. [49] investigated the crack propagation in an atomistic amorphous silica model through MD modeling with ReaxFF. Pitman and Van Duin [50] studied the clay-water interaction through the reactive MD simulation. Their proposed ReaxFF parameter set contains a large body of data, including equations of state of metal and metal-oxide phases and a series of chemical reactions involving silicon and aluminum compounds. Therefore, this parameter set is representative of all coordination states occurring during the crack of clays. However, to the best of our knowledge, no work has been done to probe the fundamental crack propagation mechanism in clay sheets using a bond-order force field considering clay-water adsorption. In this article, as a major contribution, the bond-order based ReaxFF clay force field is utilized to study the mode I and mode II cracking propagation mechanism in clay particles with water adsorption.

The remainder of this article is organized as follows. Section 2 summarizes the clay models and numerical methods, including the model setup, reactive force field parameters, and loading protocols. Section 3 presents the numerical results of the reactive MD modeling of the mode I and mode II crack propagations in the two clay minerals with water adsorption. Section 4 briefly discusses the limitation of ReaxFF for modeling cracking in clay with water adsorption and points out the future research directions, followed by a closure in Section 5.

## 2. Material models and numerical methods

### 2.1. Clay models

In this study, we adopt two clay minerals to study the water adsorption mechanisms and their potential impact on mode I and mode II crack propagation in different clay minerals. The clay minerals are pyrophyllite and Ca-montmorillonite, whose chemical formulas are $Al_2[Si_4O_{10}](OH)_2$ and $CaAl_2(Si_4O_{11})(OH)$ respectively. The difference between the two types of clay minerals is the isomorphous substitution and cation exchange capacity. Both clay minerals have a 2:1 tetrahedral-octahedral-tetrahedral (T-O-T) structure. The atomic coordinates for the unit cell of the two clay can be found in [51]. Figure 1 shows the initial molecular structure of the clay model. The MD clay model for pyrophyllite and Ca-montmorillonite consists of 720 unit cells which consist of 36 cells in the x direction, 20 cells in the y direction, and 1 cell in the z direction. Thus, the two clay platelet particles have dimensions 185.76 Å × 179.32 Å and 186.84 Å × 180.4 Å in the x-y plane, respectively. A pre-existing edge crack is then introduced to the clay sheet. The edge crack is 93 Å in length in the x-y plane with a taper angle of 5°. In the initial condition, we place a finite number of water molecules (blue in Figure 1) on the clay sheet. The water mass content in the clay-water model is 4.24% (i.e., the ratio of water molecular mass to dry clay molecular mass). Next, we introduce the ReaxFF force field and input parameters for both clay minerals.

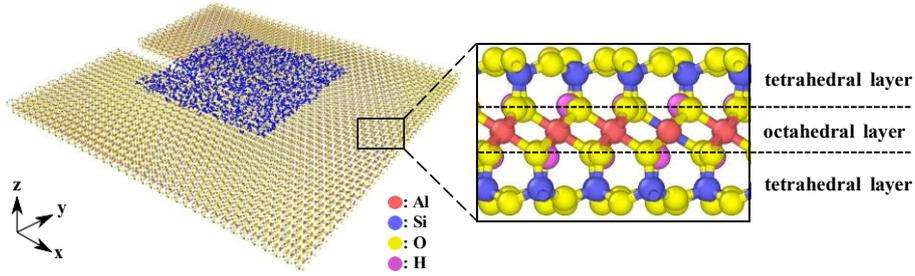

Figure 1: MD model of the initial clay-water system.

### 2.2. ReaxFF force field

The total potential energy $E_t$ of the ReaxFF force field [44] is written as

$$E_t = E_{bond} + E_{oc} + E_{uc} + E_{lp} + E_{va} + E_p + E_{ta} + E_{conj} \\ + E_{vdW} + E_{Coul}, \quad (1)$$

where $E_{bond}$ is the continuous bond energy function that describes single, double, and triple bonds, $E_{oc}$ is the over-coordination penalty energy, $E_{uc}$ is the under-coordination penalty energy, $E_{lp}$ is the lone-pair energy, $E_{va}$ is the valence angle energy, $E_p$ denotes the penalty energy, $E_{ta}$ is the torsion angle energy, $E_{conj}$ is the conjugation energy, $E_{vdW}$ is van der Waals energy, and $E_{Coul}$ is Coulomb energy. These energy terms can be categorized into two groups, bond order-dependent terms (i.e., the first 8 terms) and bond order-independent terms (the last two terms). The bond order-dependent energy terms vanish upon bond breakage. Nonbonded interactions are independent of the bond order. Both nonbonded energy terms are calculated between all atom pairs in the clay-water system. We note that ReaxFF employs the shielded potential to avoid excessive repulsion and un-physical charges between two atoms in close proximity [44].

In ReaxFF the bond order $BO'_{ij}$ between two atoms $i$ and $j$ is a function of the interatomic distance $r_{ij}$ as

$$BO'_{ij} = BO'^{\sigma}_{ij} + BO'^{\pi}_{ij} + BO'^{\pi\pi}_{ij} \\ = \exp\left[p_{bo,1}\left(\frac{r_{ij}}{r_0^{\sigma}}\right)^{p_{bo,2}}\right] + \exp\left[p_{bo,3}\left(\frac{r_{ij}}{r_0^{\pi}}\right)^{p_{bo,4}}\right] + \exp\left[p_{bo,5}\left(\frac{r_{ij}}{r_0^{\pi\pi}}\right)^{p_{bo,6}}\right], \quad (2)$$

where $BO'^{\sigma}_{ij}, BO'^{\pi}_{ij}$, and $BO'^{\pi\pi}_{ij}$, are bond orders contributed from sigma-bonds, pi-bonds, and double pi-bonds, respectively, $r_0$, $r_0$, and $r_0$ are the reference atomic distance corresponding to sigma-bonds, pi-bonds, and double pi-bonds, respectively, and $p_{bo,i}$ ($i = 1, 2, 3, 4, 5, 6$) is the scaling parameter, which are fitted from quantum mechanics calculations on small representative systems. With (2), the bond energy $E_{bond}$ can be written as

$$E_{bond} = -D_e^\sigma BO_{ij}'^\sigma \exp\left\{p_{be,1}\left[1 - \left(BO_{ij}'^\sigma\right)^{p_{be,2}}\right]\right\} - D_e^\pi BO_{ij}'^\pi - D_e^{\pi\pi} BO_{ij}'^{\pi\pi} \tag{3}$$

where $D_e$ is the dissociation energy for each bond type, and $p_{be,1}$ and $p_{be,2}$ are atomic parameters fitted from quantum mechanics calculations. The non-bonded van der Waals (vdw) interaction between atom $i$ and $j$ in ReaxFF through a distance-corrected Morse potential is written as

$$E_{vdw} = D_{ij}\left\{\exp\left[\alpha_{ij}\left(1 - \frac{f(r_{ij})}{r_{vdw}}\right)\right] - 2\exp\left[\frac{1}{2}\alpha_{ij}\left(1 - \frac{f(r_{ij})}{r_{vdw}}\right)\right]\right\} \tag{4}$$

where $D_{ij}$, $\alpha_{ij}$ and $r_{vdw}$ are parameters (see Table 2), and $f$ is the shielded interaction term as a function of the atomic distance $r_{ij}$. $f(r_{ij})$ is written as

$$f(r_{ij}) = \left[r_{ij}^{\lambda_{vdw}} + \left(\frac{1}{\gamma_w}\right)^{\lambda_{vdw}}\right]^{1/\lambda_{vdw}}, \tag{5}$$

where $\lambda_{vdw}$ and $\gamma_w$ are shielding parameters (see Table 2).

The ReaxFF parameter set for clay-water system is developed in Pitman and Van Duin [50]. Table 1 summarizes the values of parameters used to calculate the bond order and bond energy in the ReaxFF force field [50, 52]. Note that the bond order of Si-O has two components, sigma bond order and pi bond order whereas Al-O only has sigma bond order. Table 2 summarizes the parameters for the van der Waals energy

Table 1: Summary of the parameters for bond order and bond energy terms in the ReaxFF force field.

| Bond | $D_e^\sigma$ (kcal/mol) | $D_e^\pi$ (kcal/mol) | $p_{bo,1}$ | $p_{bo,2}$ | $p_{bo,3}$ | $p_{bo,4}$ |
|---|---|---|---|---|---|---|
| Si-O | 261.907 | 5.9533 | -0.1083 | 8.5924 | -0.2366 | 29.7817 |
| Al-O | 228.4876 | 0 | -0.1750 | 5.2102 | -0.3500 | 25 |
| Bond | $p_{be,1}$ | $p_{be,2}$ | $r_0^\sigma$ (Å) | $r_0^\pi$ (Å) | | |
| Si-O | -0.6223 | 10.1541 | 1.61 | 1.294 | | |
| Al-O | -0.8524 | 0.4016 | 1.5382 | - | | |

in the ReaxFF force field.

Table 2: Summary of parameters for the van der Waals energy.

| Atom $i$ | Atom $j$ | $D_{ij}$ (kcal/mol) | $\alpha_{ij}$ | $r_{vdw}$ (Å) | $\lambda_{vdw}$ | $\gamma_w$ |
|---|---|---|---|---|---|---|
| Si | O | 0.1318 | 10.5055 | 4.2555 | 1.5591 | 8.4893 |
| Al | O | 0.1526 | 9.5637 | 4.7628 | 1.5591 | 7.3490 |

Next, we show the characteristics (bond order and energies) of the Si-O and Al-O bonds in the ReaxFF force field for clay minerals. In so doing, we plot the bond order and bond and van der Waals energy curves for the Si-O bond and Al-O bond, respectively. Figure 2 plots the variation of bond order with Si-O bond length. Two components, the sigma bond order and Pi bond order, give an initial Si-O bond order of 2. The Si-O bond order vanishes as the Si-O bond length increases to 2.6 Å. Figure 3 (a) and (b) plot the bond energy $E_{bond}$ and van der Waals energy $E_{vdw}$ of the Si-O bond as a function of the bond length (the interatomic distance). Note that the energy effects related to under and over coordination are not taken into account. $E_{bond}$ remains unchanged when the bond length is less than 1 Å and then decreases when the bond length exceeds 1 Å. $E_{bond}$ reaches zero at the bond length of 2.5 Å. $E_{vdw}$ decreases monotonically with the bond length and vanishes at 2 Å.

Figure 4 plots the variation of Al-O bond order with bond length. As shown in Figure 4, the maximum bond order of Al-O is 1 due to the only sigma-bond contribution. Bond order becomes negligible at the bond length of 3 Å. Figure 5 (a) and (b) plot bond energy $E_{bond}$ and van der Waals energy $E_{vdw}$ of Al-O as a function of the bond length, respectively. The bond order energy decreases as the bond length exceeds 1 Å and becomes zero at 3 Å. $E_{vdw}$ decreases with the bond length and vanishes at the bond length of 2.5 Å. Next, we present the numerical simulations.

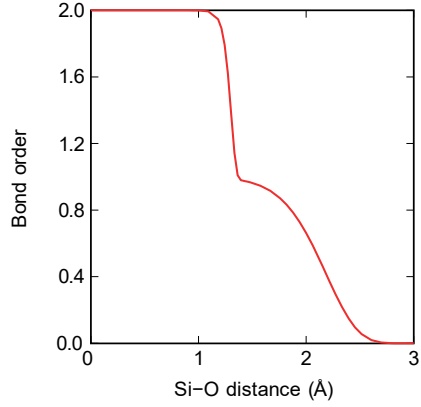

Figure 2: Variation of Si-O bond order with bond length.

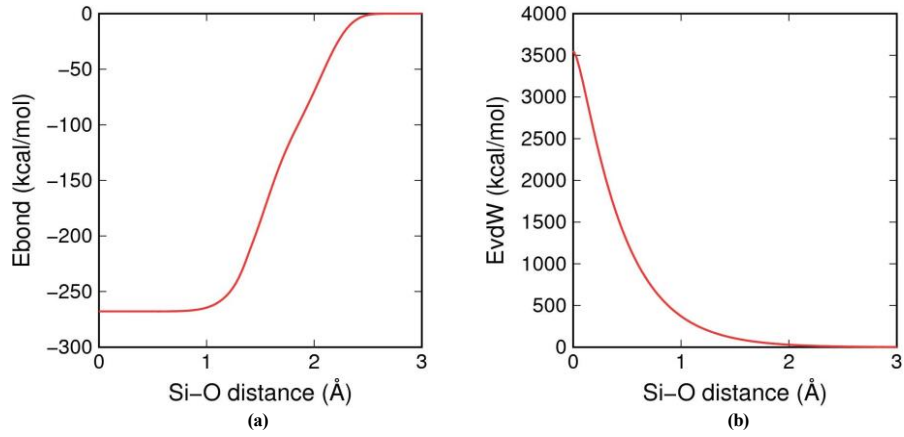

Figure 3: Bond energy and van der Waals energy of Si-O as a function of atomic distance.

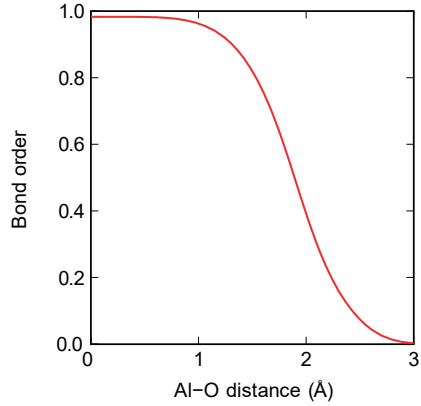

Figure 4: Variation of Al-O bond order with bond length.

*2.3. Numerical simulations*

We have implemented the ReaxFF force field in LAMMPS, a massively parallel molecular simulator [20]. Thus, all reactive MD simulations are conducted in LAMMPS. Periodic boundary conditions are applied to the clay-water system in the x, y, and z directions. The simulations are conducted in the NPT ensemble (constant number of atoms, constant pressure, and constant temperature) at 0 atm and 298 K with a time step of 0.5 fs (1 fs = $10^{-15}$ s). The temperature remains constant following the Nose-Hoover algorithm [53]. The damping parameters are 50 fs and 100 fs for temperature and pressure, respectively. The equations of motion of atoms are integrated using the Verlet velocity algorithm. Reactive MD simulations feature the flexible atomic charge. In this study, the atomic charge is adjusted at each time step using the charge equilibration method [54]. We note that MD simulations with a reactive force field are computationally more demanding than non-reactive MD. To improve computational efficiency, all simulations are conducted with 128 CPU cores on a supercomputer. The clay-water system could reach an equilibrium after 6 ns.

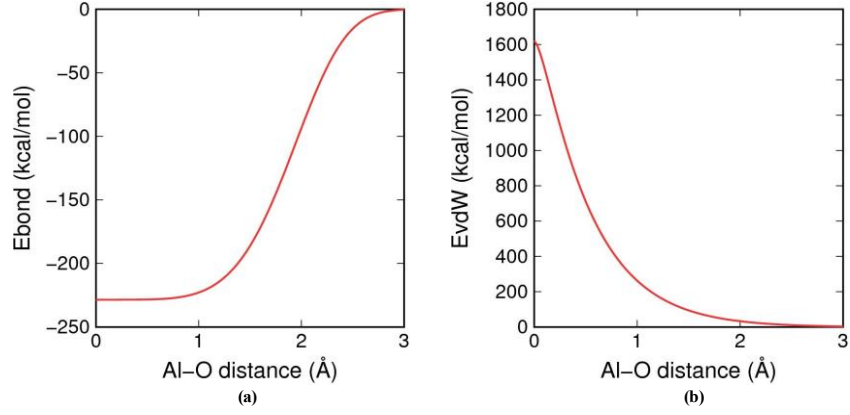

Figure 5: Variations of Bond energy and van der Waals energy of Al-O with bond length.

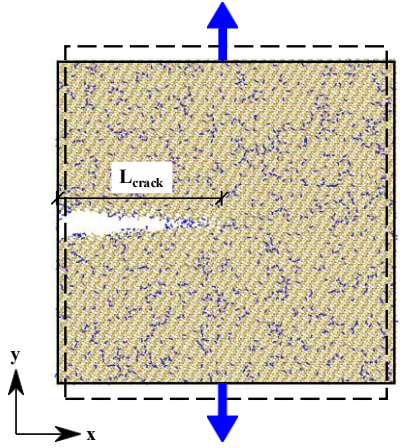

Figure 6: Schematics of mode I tension using the homogeneous deformation technique.

After the clay-water system reaches equilibrium, the tensile and shear loading are applied through the homogeneous deformation technique [20] for mode I and mode II cracking propagations, respectively. Figure 6 presents a schematic of the tensile loading through the homogeneous deformation technique. The deformation is applied by changing the size of the simulation box at a strain rate. At each time step, the simulation box dimensions are modified based on the strain, and the atomic coordinates are updated based on Newton's equations of motion and interatomic potential. It is noted that the uniform stretching of the system avoids the generation of shock waves during simulations [55]. Due to Poisson's effect, the pressure components in the x, y, and z directions are controlled independently to allow for transverse deformations. For example, as the simulation box is deformed along the y direction, we apply the free transverse pressure condition by assigning zero pressure in the x and z directions of the clay model. For post-processing, the molecular visualization tool of OVITO [56] is used to visualize and analyze the MD simulation data in this study. In what follows, we probe the mode I and mode II crack propagation mechanism in clay with water absorption from the reactive MD simulation results.

## 3. Numerical results

We present the numerical results of the MD simulation of cracking propagation in clay with water adsorption. From the results, we interpret the molecular-scale crack propagation mechanism in clay under the mode I and mode II cracks. First, we characterize the water adsorption mechanism and its potential impact on crack propagation. Second, we present the molecular-scale characteristics of the crack propagation, including crack tip blunting and bond breakage along the new crack. Third, we compute the energy release rate and stress intensity factor from our reactive MD simulations and compare our MD results with the data in the literature.

*3.1. Water adsorption mechanism*

It is known that pyrophyllite and Ca-montmorillonite have different water adsorption behavior. Figure 7 compares the distribution patterns of water molecules on the two clay surfaces. It can be seen that water molecules on pyrophyllite are dispersed in a discrete and uniform pattern. However, water molecules adsorbed to Ca-montmorillonite are in a clustered and less uniform pattern. Indeed, the locations of adsorbed water clusters on Ca-montmorillonite agree with the locations of surface cations $Ca^{2+}$.

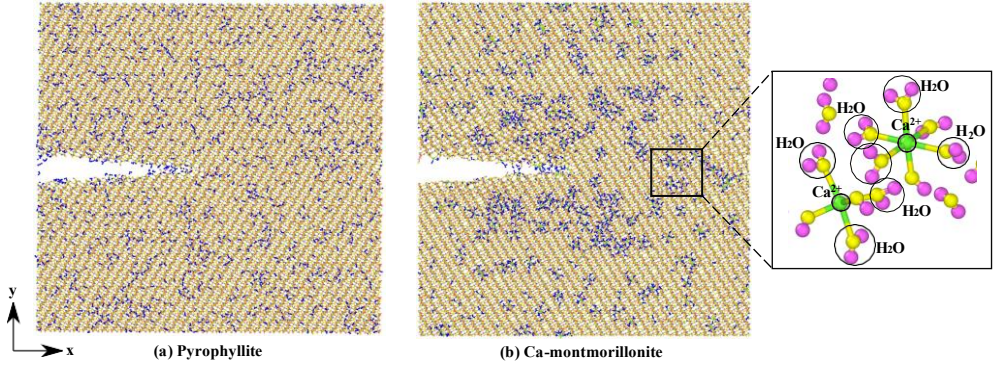

Figure 7: Distribution pattern of water molecules on (a) pyrophyllite and (b) Ca-montmorillonite.

Figure 8 shows the orientation of adsorbed water molecule on the clay surface. Water hydrogen atoms move closer to the clay surface than water oxygen atoms due to the hydrogen bond formation between clay surface oxygen and water hydrogen. This is one type of mechanism for clay-water adsorption. The hydrogen bond formation is observed in both pyrophyllite and Ca-montmorillonite. The water orientation angle $\varphi$ can be used to characterize water adsorption [57]. Here $\varphi$ is the angle between the normal to the clay surface (z-axis) and the dipole moment of water molecule, as shown in Figure 8. Figure 9 presents the evolution of water orientation angles under the mode I cracking condition. In this case, $\varphi$ is less than 22° during the crack propagation process. The relatively small angle could imply that the hydrogen bond plays a key role in the clay-water adsorption.

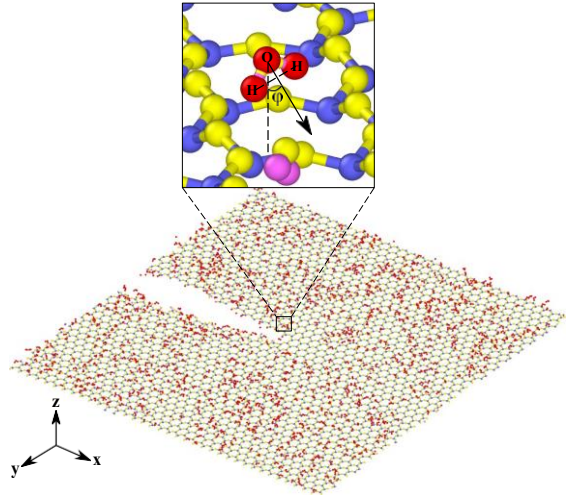

Figure 8: Schematic of the orientation angle $\varphi$ of water molecules adsorbed on the clay surface.

Next, we probe the potential impact of water adsorption on crack propagation through comparing the number density profiles of adsorbed water molecule on both clay minerals before and after the crack propagation. Figure 10 and Figure 11 show the water number density profiles for pyrophyllite and montmorillonite, respectively. The dash vertical line $z = 0$ represents the location of initial clay top surface. The peak density indicates the accumulation of water molecules above clay surface due to water adsorption. The results in Figures 10 and 11 show that the peak water number density decreases after the crack propagation. For montmorillonite, the distance from the clay top surface to the location where the peak density occurs is about 1.8 Å . For pyrophyllite, this distance is 2.3 Å . We find that the number density is not strictly equal to zero at plane $z = 0$ for montmorillonite. The reason could be that the bending occurs near the edge crack region in the clay particle. Then water molecules enter the crack region under the plane $z = 0$. This effect is more remarkable in montmorillonite. The peak number density of water molecules above the montmorillonite surface is smaller than that of pyrophyllite.

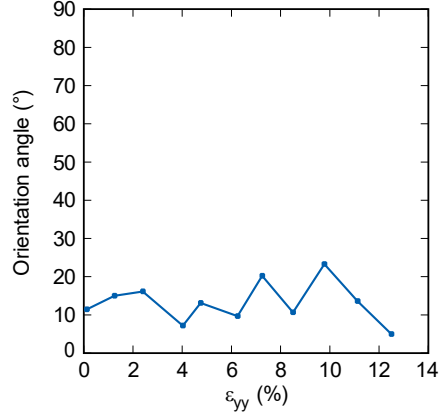

Figure 9: Variation of water orientation angles under the mode I crack condition.

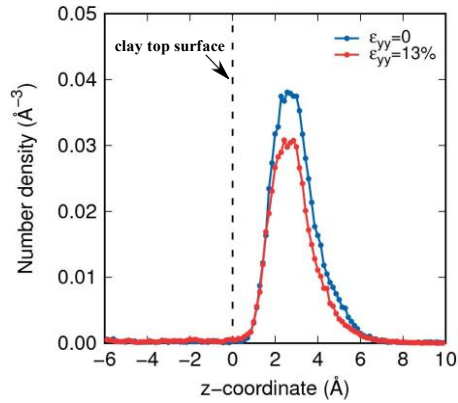

Figure 10: Number density profile of water molecule adsorbed on the pyrophyllite surface.

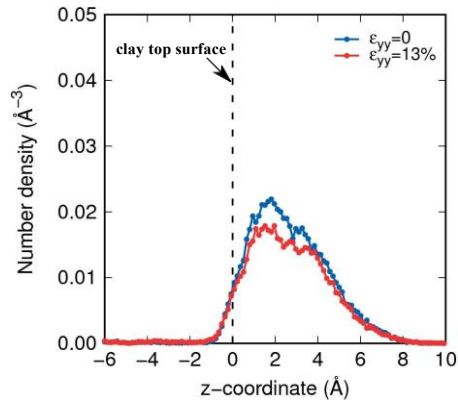

Figure 11: Number density profile of water molecules adsorbed on the montmorillonite surface.

We note that the other type of water adsorption is hydration of surface cations. This adsorption mech-anism is unique to Ca-montmorillonite due to the existence of surface cations $Ca^{2+}$. Figure 7 (b) presents cation hydration on the montmorillonite surface. It can be found from Figure 7 (b) that each $Ca^{2+}$ is fully hydrated and surrounded by water molecules. The similar arrangement of water molecules near $Ca^{2+}$ can be also found in Sposito et al. [16]. We may conclude that the difference of water distribution patterns on the two clay surfaces is attributed to surface cation hydration. In what follows, we further investigate the bending mechanism near the edge crack and silanol formation on the clay surface.

*3.1.1. Bending deformation near the edge crack*

Our MD simulation results show that the adsorbed water could cause bending deformation near edge crack. Figures 12 and 13 present the deformed edge crack region in pyrophyllite and montmorillonite, respectively. Initially, both clay sheets remain planar. During the equilibration process, the edge crack region bends upward to some extent, as indicated by the curved dash line. This effect is more remarkable for montmorillonite under the same other conditions, which may explain why the number density of water molecules at $z = 0$ is not zero in Figure 11.

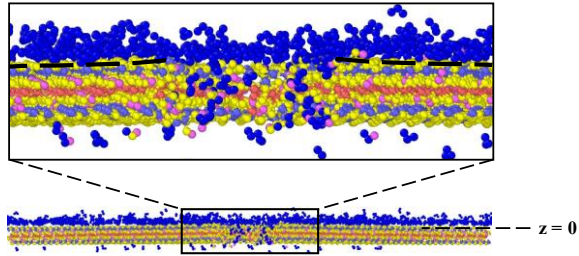

Figure 12: Bending deformation near edge crack in pyrophyllite.

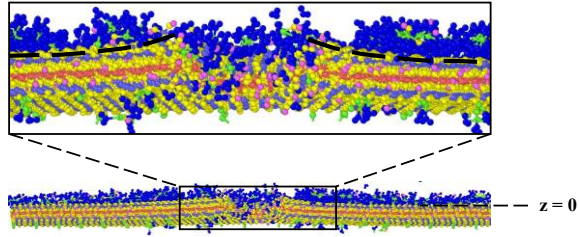

Figure 13: Bending deformation near edge crack in montmorillonite.

*3.1.2. Silanol formation*

In the cracking propagation, we observe two distinct behaviors of water molecules. Some water molecules are adsorbed to the clay surface whereas some water molecules react with clay and form new species. For instance, silanol bonds, a functional group in silicon chemistry with the connectivity Si-O-H [58, 59], are formed due to the reaction between Si in clay and the hydroxyl groups in water molecules. Clay can react with water molecules and form silanol bonds through two pathways [50] as

$$O_nSi - O^- + H_2O \rightarrow O_nSi - OH + OH^-, \tag{6}$$

$$O_nSi^+ + H_2O \rightarrow O_nSi - OH + H^+. \tag{7}$$

Figure 14 illustrates the formed silanol bonds near the crack tip. The two species generated are $(Si\text{-}O)_c\text{-}H_w$ and $Si_c\text{-}(OH)_w$. Here, the subscripts $c$ and $w$ denote that the atoms or bonds are from clay and water molecules, respectively. For example, $(Si\text{-}O)_c\text{-}H_w$ is formed by reactant $Si\text{-}O$ from clay and $H$ from water molecules.

*3.2. Mechanism of mode I and mode II crack propagation*

*3.2.1. Crack tip blunting*

We investigate the crack tip blunting behaviors by characterizing the shape change of the hexagonal Si-O cell in the tetrahedral layer. A standard hexagonal Si-O cell consists of six silicon atoms and six oxygen atoms connected through Si-O bonds. Figure 15 compares the Si-O cell in pyrophyllite at equilibrium and at the tensile strain of 11.675%. This strain is assumed as the critical strain prior to the first Si-O bond breaking. The crack tip open distance $d$ (i.e., the largest distance between two silicon atoms) and the crack tip opening angle $\alpha$ (i.e., the O-Si-O angle) are adopted to characterize the deformation of the Si-O cell in the crack propagation process. Figure 16 and 17 plot the variations of the Si-O cell open distance and the O-Si-O angle in pyrophyllite and montmorillonite under the mode I crack. The O-Si-O angle at equilibrium in pyrophyllite and montmorillonite from the MD simulation is about 113° and 108°, respectively. The values are close to the value 109.4° obtained from the X-ray measurement [60]. As shown in Figures 16 and 17 both the open distance and the O-Si-O angle increase with fluctuations and reach the maximum values before the Si-O bond breakage.

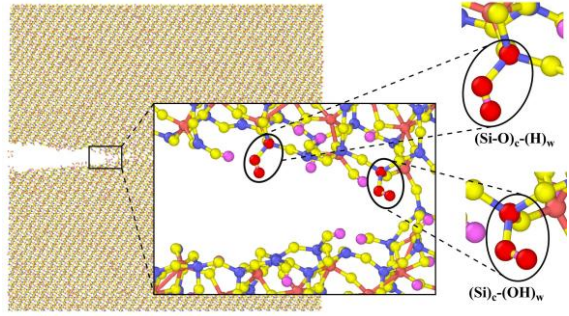

Figure 14: Two pathways for forming silanol bonds near the crack tip.

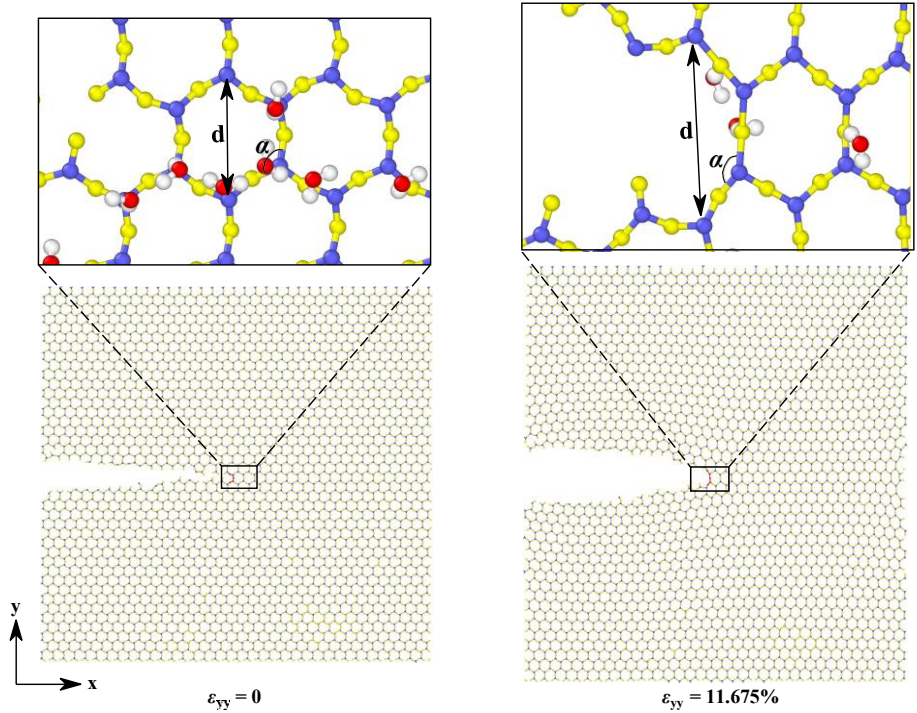

Figure 15: Configurations of Si-O cell at the crack tip at two strains under the mode I crack.

Figure 18 compares the deformation of the Si-O cell in pyrophyllite under the mode II crack. Figure 19 and 20 plot the variations of the open distance of the Si-O cell and the O-Si-O angle under the mode II crack, respectively. The maximum angle before the Si-O bond breakage is similar in both clays. However, the open distance of the Si-O cell is smaller in pyrophyllite than in montmorillonite at the same strain. For instance, the open distances of the Si-O cell in pyrophyllite and montmorillonite at the same maximum strain are 6.64 Å and 6.95 Å, respectively.

Considering the O-T-O crystalline structure of pyrophyllite, we present the microstructure evolution of the mixed "O-T (octahedral-tetrahedral) cell" that consists of the central octahedral Al-O cell and the top-layer tetrahedral Si-O cell. The deformation and cracking of the two cells will be compared in terms of Al-O and Si-O bond breaking characteristics. Figure 21 shows the atomic structure of the O-T cell at the crack tip from different view angles when the system is at equilibrium.

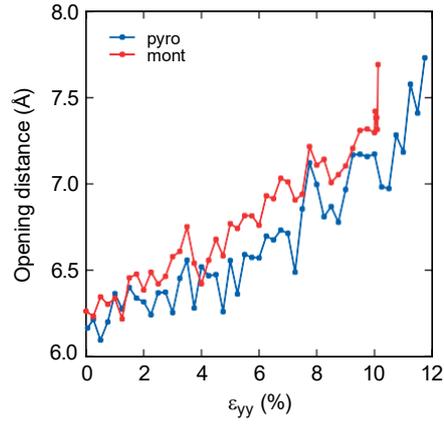

Figure 16: Variations of the open distance of the Si-O cell in pyrophyllite (pyro) and montmorillonite (mont) under the mode I crack.

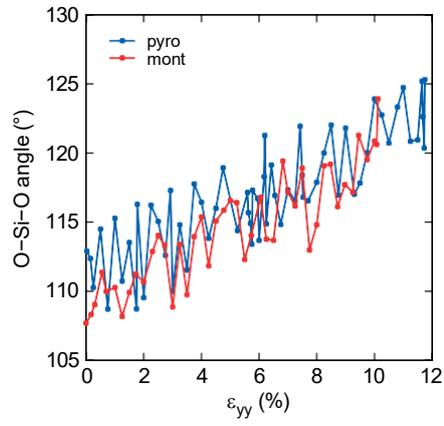

Figure 17: Variations of the O-Si-O angle in pyrophyllite and montmorillonite under the mode I crack.

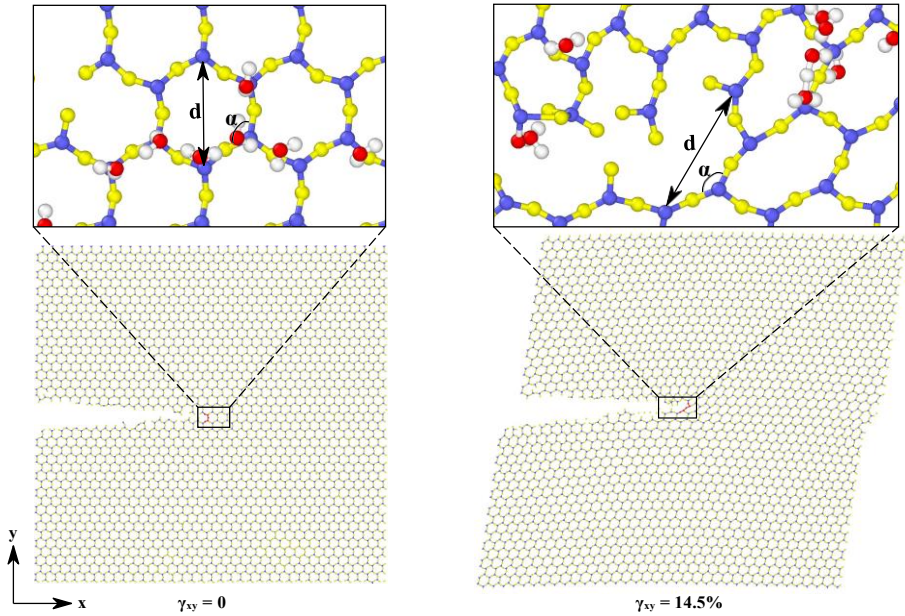

Figure 18: Deformation of the Si-O cell at the crack tip under the mode II crack.

Figure 22 plots the snapshots of deformation of the O-T cell under the mode I crack. The dash line between the two atoms indicates the bond breakage. Broken Al-O bonds are in black and broken Si-O bonds are in blue. As the tensile strain reaches 9.1%, the Al-O bond breakage occurs while the Si-O cell still remains intact. The breakage of the Si-O bond in this cell is not observed until $\varepsilon_{yy} = 11.725\%$. The bond breakage in the Si-O cell indicates the occurrence of the crack propagation. Figure 23 presents the deformation of the O-T cell under the mode II crack. The first bond breakages in the octahedral Al-O cell and the tetrahedral Si-O cell occur at the shear strains of 12.97% and 14.44%, respectively. Despite the earlier breakage, the broken Al-O bond does not immediately result in the crack growth in that the the Si-O cell can still sustain the external load. Indeed, the bond breakage in the Si-O cell is directly associated with the crack growth. Thus, we may conclude that the tetrahedral Si-O cell is primarily in charge of the crack tip stability.

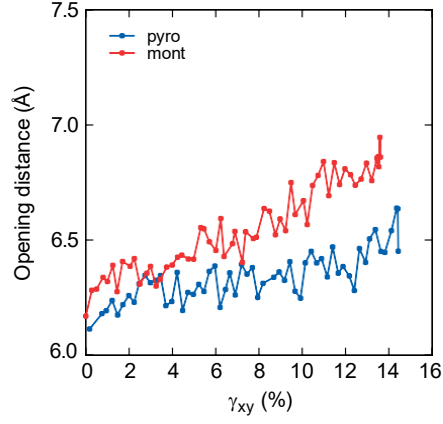

Figure 19: Variations of the open distance of the Si-O cell in pyrophyllite and montmorillonite under the mode II crack.

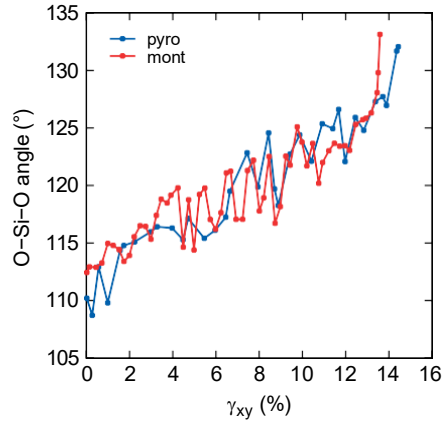

Figure 20: Variations of O-Si-O angle in pyrophyllite and montmorillonite under the mode II crack.

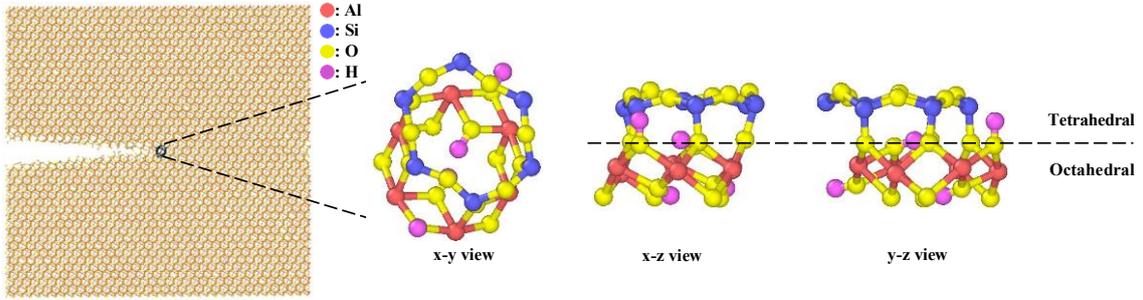

Figure 21: Atomic structure of the O-T cell.

In what follows, we characterize the crack tip blunting by calculating the radius of the crack tip curvature ($\rho$). Figure 24 presents the schematic of $\rho$ based on a second-order polynomial curve fitting. In this case, the radius of curvature can be written as

$$\rho = \frac{1}{2a} \tag{8}$$

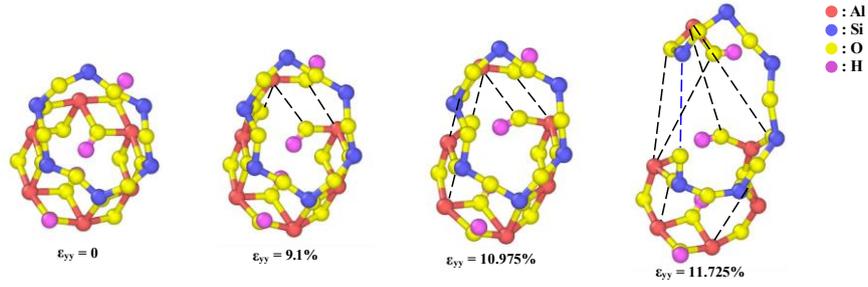

Figure 22: Snapshots of the deformation of the O-T cell under the mode I crack.

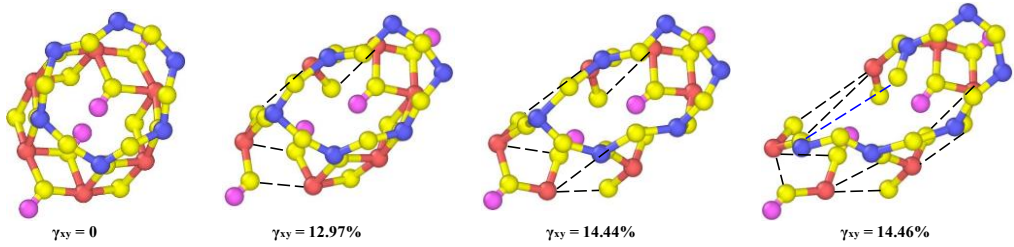

Figure 23: Snapshots of the deformation of the O-T cell under the mode II crack.

where $a$ is the coefficient in the second-order polynomial function. Figure 25 shows the variations of the radii of the crack tip in both clays under the mode I crack. In Figure 25, the two vertical dash lines denote the critical strains at which the crack starts to propagate. In the crack tip blunting process, the radius of curvature first experiences a stable stage. Then the radius of curvature gradually increases to the maximum value followed by a sharp drop upon the crack propagation. The crack blunting before its propagation is remarkable because about six-time increases in the radius of curvature are observed for both clay models. We note that the critical strain corresponding to the abrupt decrease in $\rho$ is consistent with the abrupt decline in the stress-strain curve. Figures 26 and 27 plot the stress-strain curves of pyrophyllite and montmorillonite under the mode I and mode II cracks, respectively. Both pyrophyllite and montmorillonite have a similar peak tensile stress under the mode I crack. However, under the mode II crack, pyrophyllite has a larger peak shear stress than montmorillonite. Next, we analyze the bond breakage mechanism under both crack modes.

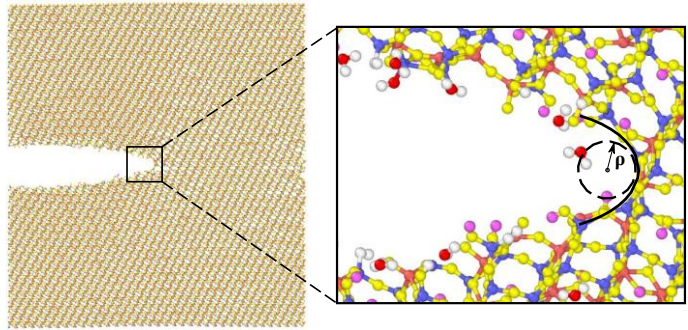

Figure 24: Schematic of the radius of the crack tip curvature.

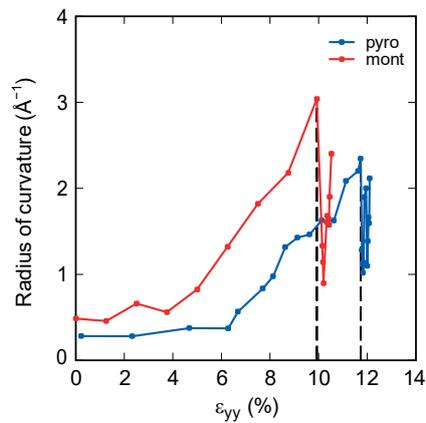

Figure 25: Variation of the radius of the crack tip curvature in pyrophyllite and montmorillonite under the mode I crack.

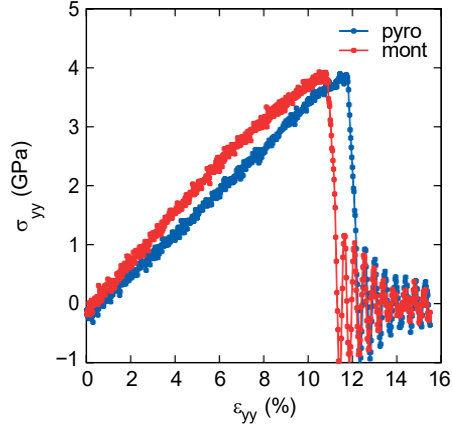

Figure 26: Stress strain curves of pyrophyllite and montmorillonite under the mode I crack.

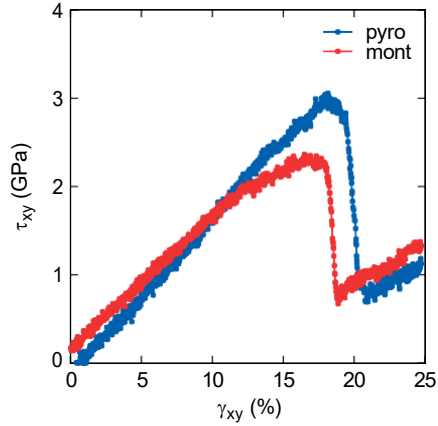

Figure 27: Stress strain curves of pyrophyllite and montmorillonite under the mode II crack.

*3.2.2. Bond breakage characteristics*

In this part, we examine the individual bond breakage behavior of Al-O and Si-O bonds, as well as the variations of broken bond number and crack length under both crack modes. Figure 28 plots the mode I crack propagation in the pyrophyllite particle. It can be found that crack propagates forward in a straight way and ultimately breaks into two halves. The process of crack propagation takes about 4.75 ps and the corresponding strain increment is 0.475%, which may imply a brittle fracturing process. Figure 29 plots the mode II crack propagation in pyrophyllite. Different from the mode I crack propagation, the results in Figure 29 show the clay particle does not fall apart completely under the mode II crack, which may imply a ductile fracturing process. Figures 30 and 31 plot the snapshots of the mode I and mode II crack propagations respectively in montmorillonite. The results in Figures 30 and 31 show that the mode I crack propagation in the montmorillonite particle is brittle while the mode II crack propagation could be ductile. We note that this conclusion needs further studies.

Figure 32 plots the bond length variations of the Al-O and Si-O bonds under the mode I and mode II cracks. Based on our numerical simulations, the equilibrium lengths of the Si-O and Al-O bonds are about 1.6 Å and 1.95 Å, respectively. The numerical results are consistent with the equilibrium bond lengths obtained from the radial distribution function, i.e., 1.63 Å and 1.99 Å as shown in Figure 33 and 34. Our numerical results show that the maximum length of Si-O bonds before breakage is increased by 29%, and the length of the Al-O bonds are increased by 16.5%.

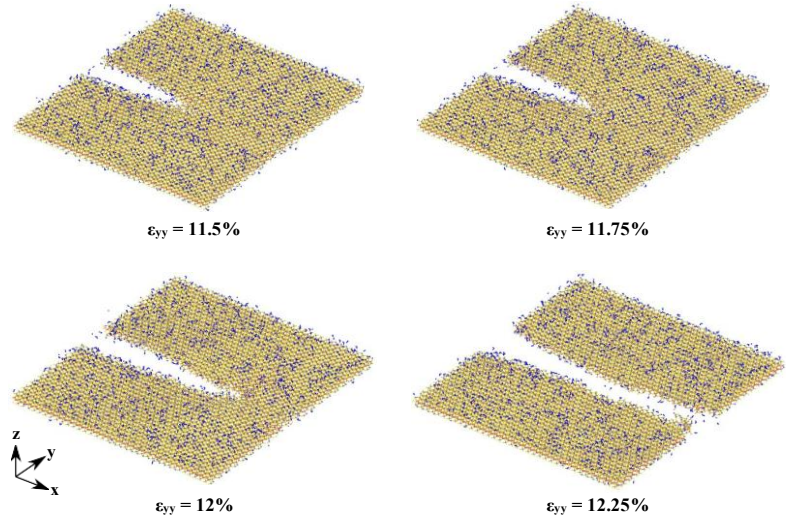

Figure 28: Snapshots of the mode I crack propagation in pyrophyllite.

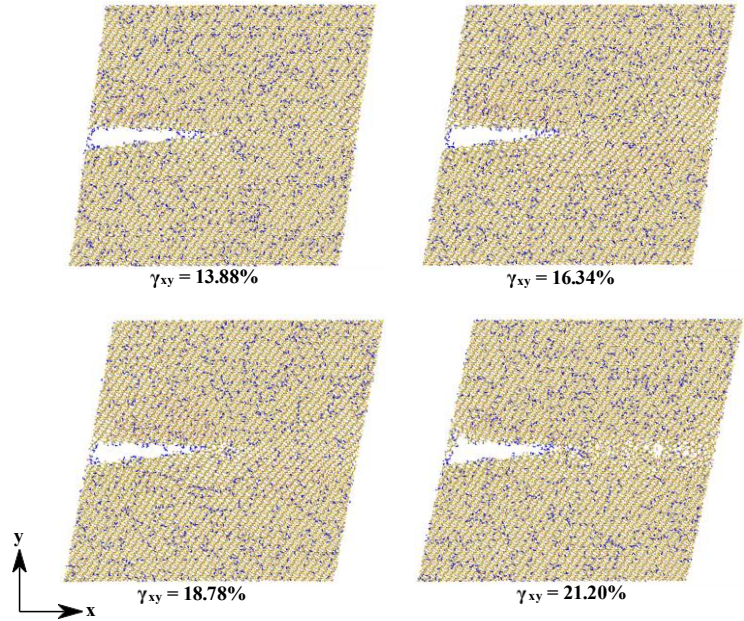

Figure 29: Snapshots of the mode II crack propagation in pyrophyllite.

The number of broken bonds can be used to characterize the crack propagation process at the atomic scale. Figure 35 plots the variation of the broken Al-O bond number in montmorillonite under mode I and mode II cracks. The sudden drop of the curve of the broken bond number after its peak value indicates the crack propagation. Figure 36 plots the variation of the broken Si-O bond number in montmorillonite under the mode I crack. The results show that the Si-O bond breakage differs from that of the Al-O bond. For instance, there is no Si-O bond breakage until the maximum tensile strain is reached. Figures 37 and 38 plot variations of the broken Al-O and Si-O bonds in pyrophyllite, respectively. The results in Figures 37 and 38 show that the variations of broken bond numbers in pyrophyllite follow the similar trends as in montmorillonite.

Figures 39 and 40 plot the crack length change with respect to the strain under the mode I and mode II cracks, respectively. It may be implied from Figures 39 and 40 that the mode I crack propagation in both clay particles are brittle while the mode II crack propagation is ductile. Meanwhile, the results in Figures 39 and 40 show that the full propagation of both mode cracks in pyrophyllite may require a larger strain or load under the same other conditions.

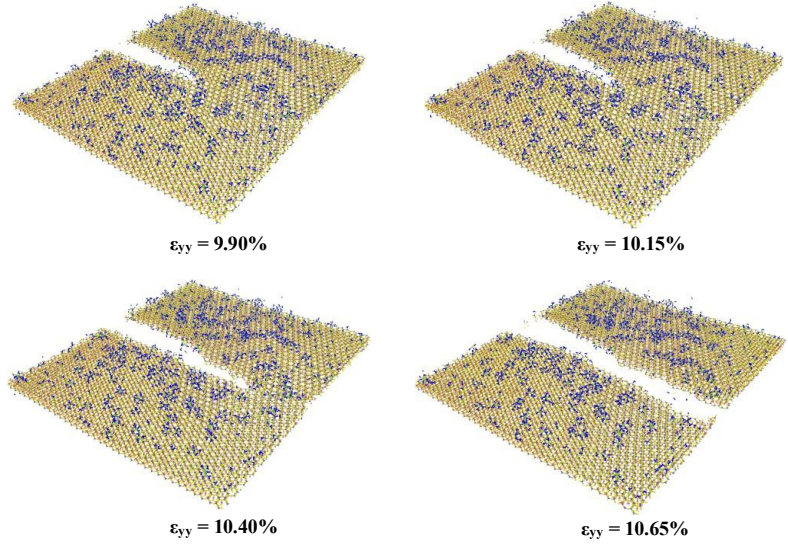

Figure 30: Snapshots of the mode II crack propagation in montmorillonite.

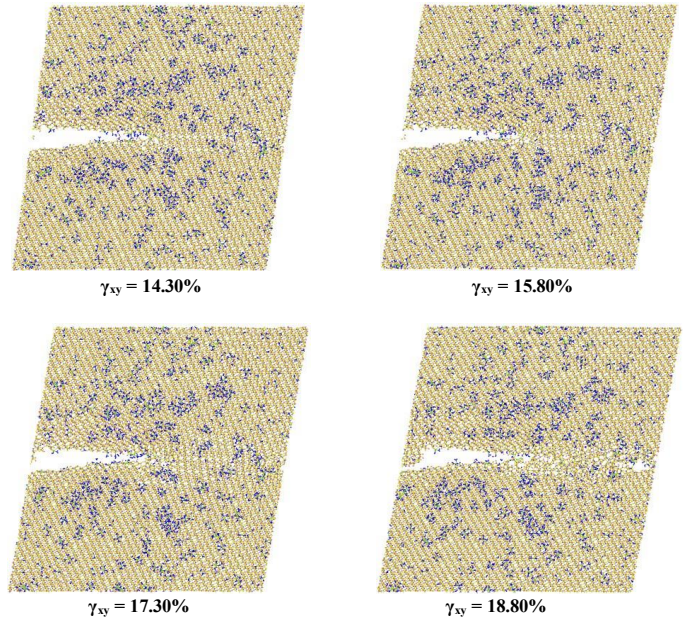

Figure 31: Snapshots of the mode II crack propagation in montmorillonite.

Next, the reactive MD simulation results are used to calculate the crack energy release rate and the stress intensity factor.

### 3.3. Crack energy release rate

Crack energy release rate $G$ is a fundamental variable in fracture mechanics at the continuum scale. Crack energy release rate is defined as the dissipated energy during fracture per unit crack surface area. In the linear elastic fracture mechanics, the crack propagates when the energy release rate reaches a critical value $G_C$. The J-integral is a method for calculating the energy release rate [61]. The integration of stress $\sigma_{yy}$ with respect to strain $\varepsilon_{yy}$ provides an estimate of the variation of the free energy per unit volume. At the end of the simulation, the mechanical energy is assumed to be completely released by the creation of new crack surfaces. Under the mode I crack, the critical energy release rate $G_{IC}$ can be written as

$$G_{IC} = \frac{V}{2(L_c - L_0) \times t} \int_0^{\varepsilon_f} \sigma_{yy} d\varepsilon_{yy} \qquad (9)$$

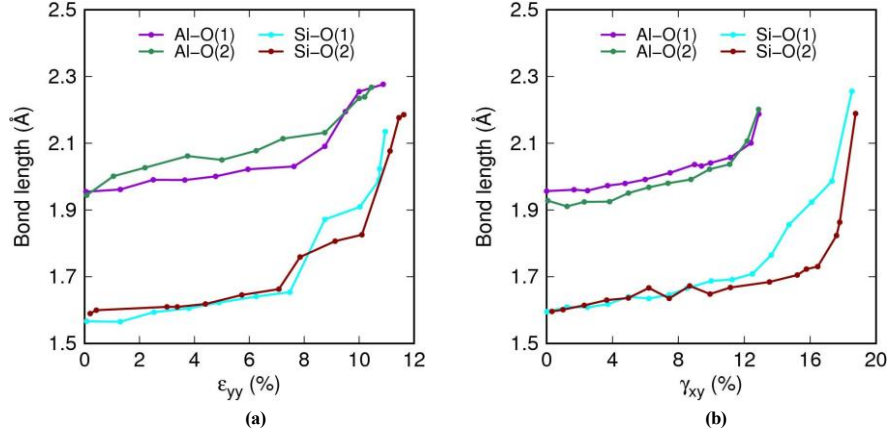

Figure 32: Variation of the bond length in pyrophyllite under the (a) mode I and (b) mode II cracks.

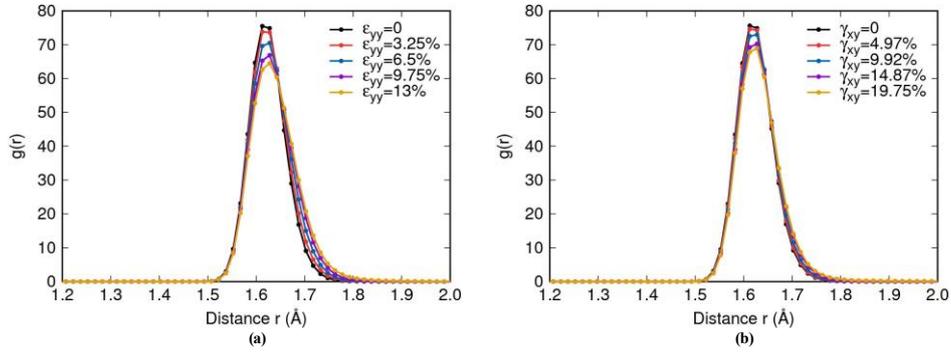

Figure 33: Radial distribution function of the Si-O bond in pyrophyllite under the (a) mode I and (b) mode II cracks.

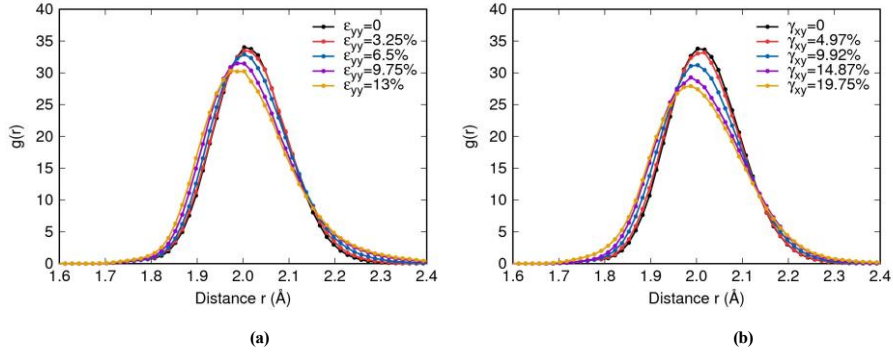

Figure 34: Radial distribution function of the Al-O bond in pyrophyllite under the (a) mode I and (b) mode II cracks.

where $V$ is the system volume, $L_c$ is the final crack length, $L_0$ is the initial crack length, $t$ is the crack thickness, $\varepsilon_f$ is the final strain, and the factor 2 denotes the two surfaces of the crack. Figure 41 presents the accumulated strain energy in pyrophyllite and montmorillonite under the mode I crack. From our reactive MD simulations and through equation (9), the range of critical energy release rate for pyrophyllite is from 3.9 to 4.8 N/m and for montmorillonite the range is from 3.7 to 5.1 N/m. Our numerical results are consistent with the experimental results in the literature (e.g., [62, 63]). For instance, the critical energy release rate of Barcelona clay is 4.6 N/m through experiments reported in [62]. The energy release rate of a clayey soil varies from 4.3 N/m to 6 N/m as shown in [63]. Next, we will present the our numerical results of the stress intensity factor.

### 3.4. Stress intensity factor

In this part, we compute the stress intensity factor from our reactive MD simulation results. For a semi-infinite platelet clay particle with an edge crack, the stress intensity factor for the mode I crack can

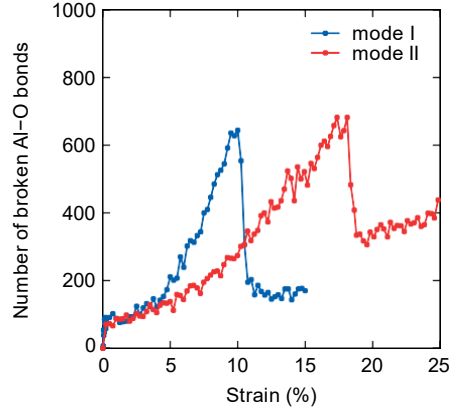

Figure 35: Number of the broken Al-O bonds in montmorillonite under the mode I and mode II cracks.

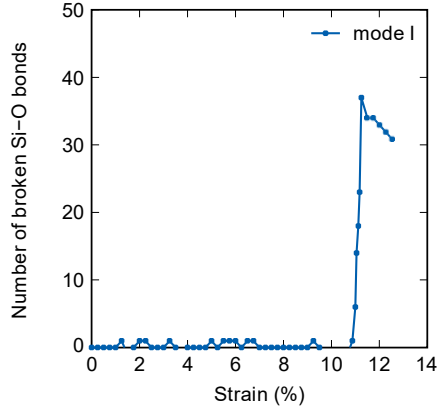

Figure 36: Number of the broken Si-O bonds in montmorillonite under the mode I crack.

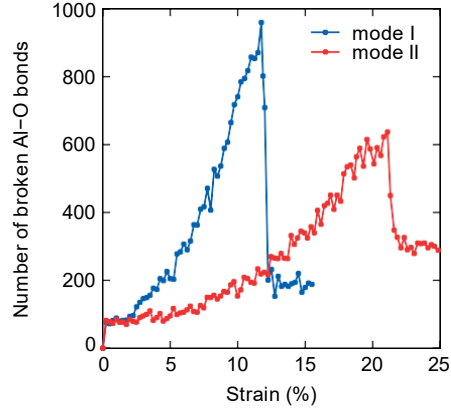

Figure 37: Number of the broken Al-O bonds in pyrophyllite under the mode I and mode II cracks.

be written as
$$K_I = 1.12\sigma\sqrt{\pi L_{cp}}, \qquad (10)$$
where $\sigma$ is the tensile stress and $L_{cp}$ is the propagated crack length. Similarly, the stress intensity factor for mode II cracks can be obtained by replacing tensile stress $\sigma$ in Equation (10) by the shear stress $\tau$ as
$$K_{II} = 1.12\tau\sqrt{\pi L_{cp}}. \qquad (11)$$

Figures 42 and 43 show the variations of the stress intensity factor in pyrophyllite and montmorillonite, respectively. From equation (10), the critical fracture toughness $K_{IC}$ for the mode I crack of pyrophyllite is 0.77 MPa·m$^{1/2}$ and the value for montmorillonite is 0.72 MPa·m$^{1/2}$. From equation (11), the critical mode II stress intensity factors for pyrophyllite and montmorillonite are 0.67 MPa·m$^{1/2}$ and 0.47 MPa·m$^{1/2}$, respectively. We compare the MD simulation results against experimental data in the literature. In what follows, we present some results in the literature to validate our numerical results. The ring tests on CL soils in [64] showed that the values of $K_{IC}$ ranges from 0.12 to 0.2 MPa·m$^{1/2}$. In [65] the authors reported that the fracture toughness of phyllosilicates mica is 0.2 MPa·m$^{1/2}$. The nanoindentation tests in [66] showed that the fracture toughness for Opalinus clay could vary between 0.316 and 0.698 MPa·m$^{1/2}$. It is noted that the results from our reactive MD simulation considering water adsorption are consistent with the data reported in the aforementioned studies.

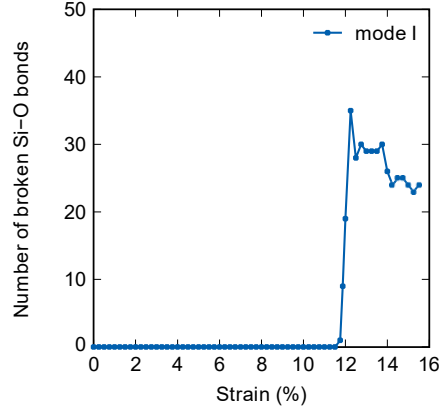

Figure 38: Number of the broken Si-O bonds in pyrophyllite under the mode I crack.

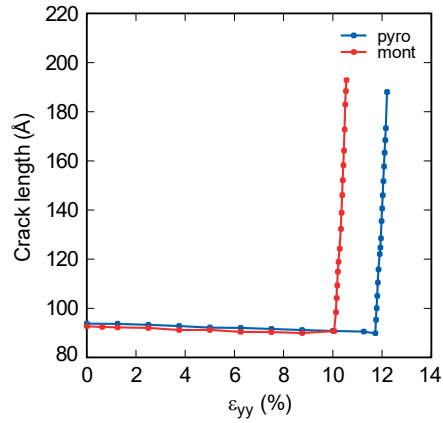

Figure 39: Evolution of the crack length in pyrophyllite and montmorillonite under the mode I crack.

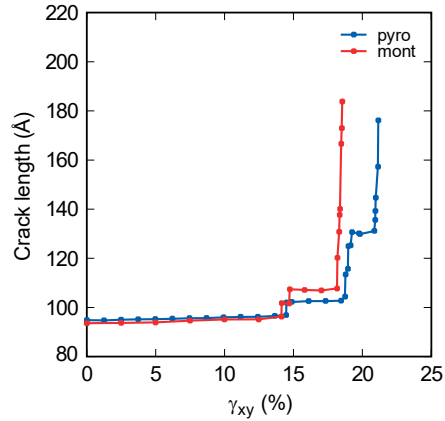

Figure 40: Evolution of the crack length in pyrophyllite and montmorillonite under the mode II crack.

We also calculate the crack process zone radius $r_p$ at the crack tip from the stress intensity factor $K_1$

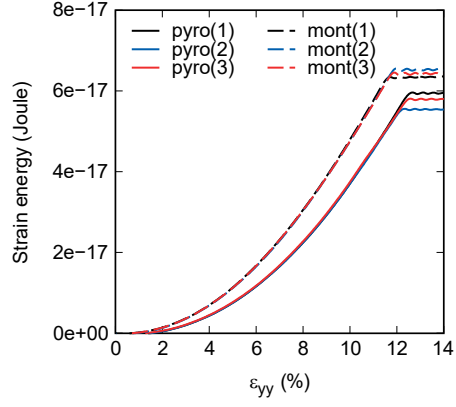

Figure 41: Strain energy in pyrophyllite and montmorillonite under the mode I crack.

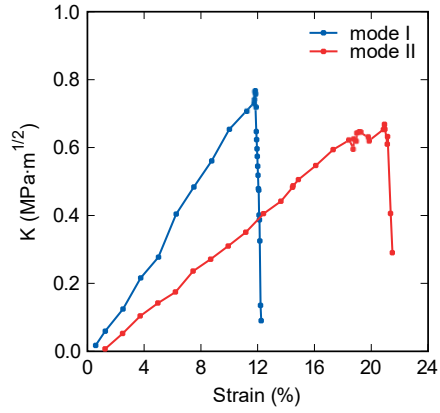

Figure 42: Variations of the stress intensity factor of pyrophyllite under the mode I and mode II cracks.

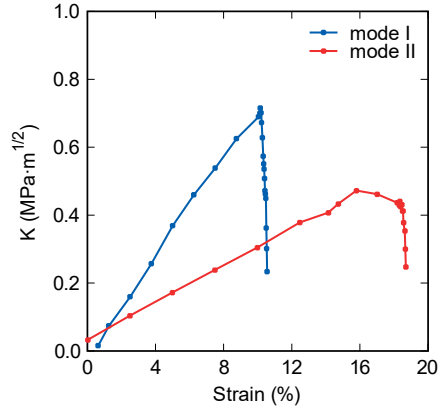

Figure 43: Variations of the stress intensity factor of montmorillonite under the mode I and mode II cracks.

and the yield stress $\sigma_y$ [67] as

$$r_p = \frac{1}{6\pi}\left(\frac{K_I}{\sigma_y}\right)^2. \tag{12}$$

In this study that the ultimate stress is the yield stress. Figure 44 plots variations of the crack process zone radius at the crack tip under the mode I crack. The maximum radius of the crack process zone at the crack tip for pyrophyllite is about 24.2 Å and the value for montmorillonite is about 20 Å from our MD simulation results. In the next section, we will briefly discuss the phenomenon and reasons of the artificial bond formation and crack healing observed in our reactive MD simulations in this study.

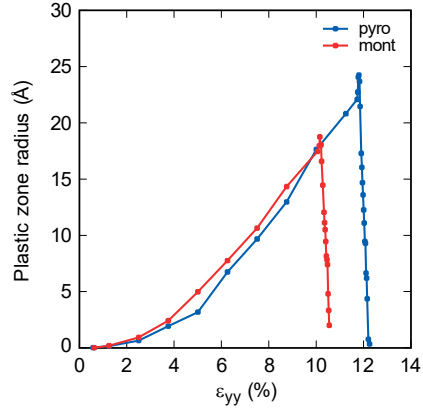

Figure 44: Variations of the crack process zone radius at the crack tip in pyrophyllite and montmorillonite under the mode I crack.

## 4. Discussion

Our numerical results show that artificial bond formation and crack healing occur during the equilibrium and crack propagation. In this part, we analyze the artificial bond formation in our reactive MD simulations and interpret the mechanism related to the adopted reactive force field.

*4.1. Artificial bond formation during equilibration*

During equilibration, it is found that artificial Al-O bonds form across the crack surfaces. Figure 45 compares the crack region before and after the artificial bond formation near crack tip. Atom 1 and 4 are aluminum and they are connected to oxygen atoms 2 and 5, respectively. Thus, the Al-O-Al type bend angles are formed, i.e., 1-2-3, and 4-5-6. Figure 46 describes the bond breakage process of the artificial Al-O bonds formed during equilibration. The first artificial Al-O bond breaks at $\varepsilon_{yy} = 6.4\%$. This artificial bond breakage occurs earlier than the real Al-O bond breakage at the crack tip which occurs at $\varepsilon_{yy} = 8.525\%$. The second artificial Al-O bond breaks at $\varepsilon_{yy} = 7.7\%$. We note that the breakage of the two artificial Al-O bonds does not contribute to the crack propagation. Figure 47 shows the variation of the open distance $d$ and open angle $\alpha$ of the Al-O cell containing the artificial bonds. Prior to bond breakage, the open distance reaches 6.5 Å and the O-Al-O angle increases to 141° from its equilibrium value of 89°.

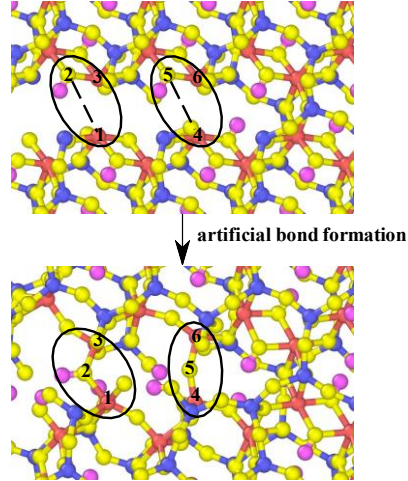

Figure 45: Artificial bond formation near the crack tip in pyrophyllite during equilibration.

*4.2. Artificial bond formation under the mode II crack*

Unlike the mode I crack propagation in which broken bonds cannot be recovered, the dynamic bond formation at the crack tip is observed in mode II crack propagation. Figure 48 illustrates the bond breakage and reformation of the Si-O bond under the mode II crack. Atom 1 and atom 2 are originally connected within the Si-O bond, and atom 3 is an isolated atom. As the shear deformation proceeds, the original

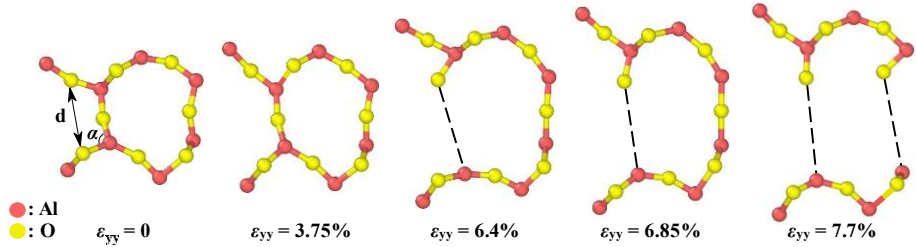

Figure 46: Bond breakage process of the artificial Al-O bonds in pyrophyllite.

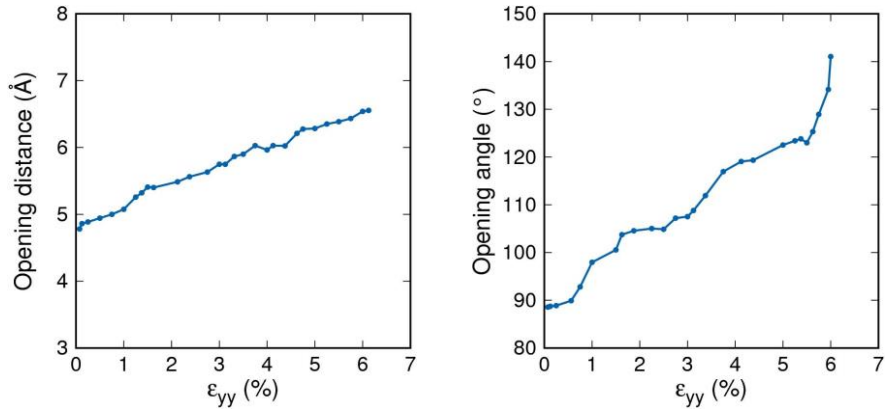

Figure 47: Variation of the open distance and O-Al-O angle of the Al-O cell at the crack tip under the mode I crack.

Si-O bond between atom 1 and atom 2 breaks. A covalent bond is then formed between atom 2 and atom 3, which does not exist initially. Consequently, the artificial bond formation results in the nearly constant number of Si-O bonds in the clay particle under the mode II crack.

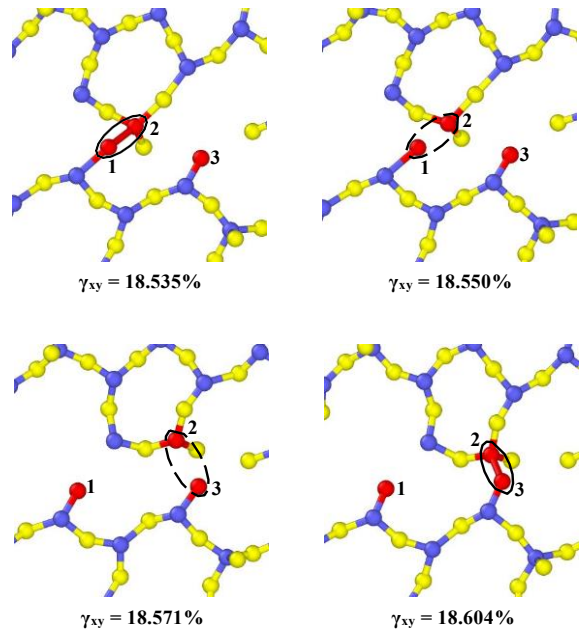

Figure 48: Breakage and formation of the Si-O bond under the mode II crack.

The artificial bond formation during the equilibrium and mode II crack is due to the nature of ReaxFF, a chemical-reaction bond order force field. Further improvement can be made by developing a deformation-driven bond order force field for more accurate modeling of crack propagation in clays with water adsorption.

## 5. Closure

In this article, we have investigated the fundamental mechanisms of mode I and mode II crack propagations in two clay minerals (pyrophyllite and Ca-montmorillonite) with water adsorption through reactive molecular dynamics simulations. Clay water adsorption is explicitly considered by adding water molecules to the clay surface. The relatively small orientating angle of the water molecule indicates the formation of hydrogen bonds throughout the crack propagation process. The peak number density of adsorbed water decreases with the increasing strains. Our numerical results show that the crack tip first gets blunted with a significant increase in the radius of the curvature of the crack tip and a slight change in crack length. The crack tip blunting is studied by tracking the evolution of crack tip opening distance, and O-Si-O angle in the tetrahedral Si-O cell under the mode I and mode II cracks. The bond breakage behaviors between Al-O and Si-O are compared, which shows that Si-O bond breaking is primarily responsible for crack propagation. The critical stress intensity factor and critical energy release rate are determined from our reactive MD simulation results and are validated against the data in the literature. The range of the critical stress intensity factor is [0.47, 0.77] MPa·m$^{1/2}$. The range of the energy release rate is [3.5, 5.1] N/m.

**Acknowledgment**

This work has been supported by the US National Science Foundation under contract numbers 1659932 and 1944009. The support is greatly acknowledged. The authors are grateful to Professor Adri van Duin for providing the input parameters of the ReaxFF force field for the two clay minerals in this study.

**References**

[1] Terzaghi K, Peck RB, Mesri G. Soil mechanics in engineering practice. John Wiley & Sons; 1996.
[2] Lambe TW, Whitman RV. Soil mechanics; vol. 10. John Wiley & Sons; 1991.
[3] Mitchell JK, Soga K, et al. Fundamentals of soil behavior; vol. 3. John Wiley & Sons New York; 2005.
[4] Lu N, Godt JW. Hillslope hydrology and stability. Cambridge University Press; 2013.
[5] Gens A. Soil–environment interactions in geotechnical engineering. Géotechnique 2010;60(1):3–74.
[6] Alonso EE. Triggering and motion of landslides. Géotechnique 2021;71(1):3–59.
[7] Menon S, Song X. Computational multiphase periporomechanics for unguided cracking in unsaturated porous media. International Journal for Numerical Methods in Engineering 2022;123(12):2837–71.
[8] Likos WJ, Song X, Xiao M, Cerato A, Lu N. Fundamental challenges in unsaturated soil mechanics. In: Geotechnical fundamentals for addressing new world challenges. Springer; 2019, p. 209–36.
[9] Tuller M, Or D, Dudley LM. Adsorption and capillary condensation in porous media: Liquid retention and interfacial configurations in angular pores. Water resources research 1999;35(7):1949–64.
[10] Lu N, Khorshidi M. Mechanisms for soil-water retention and hysteresis at high suction range. Journal of Geotechnical and Geoenvironmental Engineering 2015;141(8):04015032.
[11] Schoonheydt R, Johnston C. Surface and interface chemistry of clay minerals. Developments in clay science 2006;1:87–113.
[12] Martin RT. Adsorbed water on clay: A review. Clays and clay minerals 1962;:28–70.
[13] Marry V, Turq P. Microscopic simulations of interlayer structure and dynamics in bihydrated heteroionic montmorillonites. The Journal of Physical Chemistry B 2003;107(8):1832–9.
[14] Tunega D, Gerzabek MH, Lischka H. Ab initio molecular dynamics study of a monomolecular water layer on octahedral and tetrahedral kaolinite surfaces. The Journal of Physical Chemistry B 2004;108(19):5930–6.
[15] Zhang Z, Song X. Nonequilibrium molecular dynamics (nemd) modeling of nanoscale hydrodynamics of clay-water system at elevated temperature. International Journal for Numerical and Analytical Methods in Geomechanics 2022;.
[16] Sposito G, Skipper NT, Sutton R, Park Sh, Soper AK, Greathouse JA. Surface geochemistry of the clay minerals. Proceedings of the National Academy of Sciences 1999;96(7):3358–64.
[17] Lu N. Unsaturated soil mechanics: Fundamental challenges, breakthroughs, and opportunities. Journal of Geotechnical and Geoenvironmental Engineering 2020;146(5):02520001.
[18] Allen MP, Tildesley DJ. Computer simulation of liquids. Oxford university press; 2017.
[19] Frenkel D, Smit B. Understanding molecular simulation: from algorithms to applications; vol. 1. Elsevier; 2001.
[20] Plimpton S. Fast parallel algorithms for short-range molecular dynamics. Journal of computational physics 1995;117(1):1–19.
[21] Song X, Wang MC. Molecular dynamics modeling of a partially saturated clay-water system at finite temperature. International Journal for Numerical and Analytical Methods in Geomechanics 2019;43(13):2129–46.
[22] Cygan RT, Liang JJ, Kalinichev AG. Molecular models of hydroxide, oxyhydroxide, and clay phases and the development of a general force field. The Journal of Physical Chemistry B 2004;108(4):1255–66.
[23] Song X, Wang MC, Zhang K. Molecular dynamics modeling of unsaturated clay-water systems at elevated temperature. In: The 7th International Conference on Unsaturated Soils 2018 (UNSAT2018)-Ng, Leung, Chiu & Zhou (Eds) The Hong Kong University of Science and Technology, ISBN 978-988-78037-3-7. 2018,.
[24] Song X, Zhang Z. Determination of clay-water contact angle via molecular dynamics and deep-learning enhanced methods. Acta Geotechnica 2021;:1–15.
[25] Zhang Z, Song X. Characterizing the impact of temperature on clay-water contact angle in geomaterials during extreme events by deep learning enhanced method. In: Geo-Extreme 2021. 2021, p. 160–8.
[26] Zhang Z, Song X. Nanoscale soil-water retention curve of unsaturated clay via md and machine learning. arXiv preprint arXiv:220104949 2021;.

[27] Sato H, Ono K, Johnston CT, Yamagishi A. First-principles studies on the elastic constants of a 1: 1 layered kaolinite mineral. American Mineralogist 2005;90(11-12):1824–6.
[28] Militzer B, Wenk HR, Stackhouse S, Stixrude L. First-principles calculation of the elastic moduli of sheet silicates and their application to shale anisotropy. American Mineralogist 2011;96(1):125–37.
[29] Zartman GD, Liu H, Akdim B, Pachter R, Heinz H. Nanoscale tensile, shear, and failure properties of layered silicates as a function of cation density and stress. The Journal of Physical Chemistry C 2010;114(4):1763–72.
[30] Teich-McGoldrick SL, Greathouse JA, Cygan RT. Molecular dynamics simulations of structural and mechanical properties of muscovite: pressure and temperature effects. The Journal of Physical Chemistry C 2012;116(28):15099–107.
[31] Hantal G, Brochard L, Laubie H, Ebrahimi D, Pellenq RJM, Ulm FJ, et al. Atomic-scale modelling of elastic and failure properties of clays. Molecular Physics 2014;112(9-10):1294–305.
[32] Schmidt SR, Katti DR, Ghosh P, Katti KS. Evolution of mechanical response of sodium montmorillonite interlayer with increasing hydration by molecular dynamics. Langmuir 2005;21(17):8069–76.
[33] Katti DR, Schmidt SR, Ghosh P, Katti KS. Molecular modeling of the mechanical behavior and interactions in dry and slightly hydrated sodium montmorillonite interlayer. Canadian Geotechnical Journal 2007;44(4):425–35.
[34] Ebrahimi D, Pellenq RJM, Whittle AJ. Nanoscale elastic properties of montmorillonite upon water adsorption. Langmuir 2012;28(49):16855–63.
[35] Carrier B, Vandamme M, Pellenq RJM, Van Damme H. Elastic properties of swelling clay particles at finite temperature upon hydration. The Journal of Physical Chemistry C 2014;118(17):8933–43.
[36] Qomi MA, Ebrahimi D, Bauchy M, Pellenq R, Ulm F. Methodology for estimation of nanoscale hardness via atomistic simulations. Journal of Nanomechanics and Micromechanics 2017;7(4):04017011.
[37] Jia X, Hao Y, Li P, Zhang X, Lu D. Nanoscale deformation and crack processes of kaolinite under water impact using molecular dynamics simulations. Applied Clay Science 2021;206:106071.
[38] Zhang Z, Song X. Md modeling of cracks in clay at the nanoscale. arXiv preprint arXiv:220105969 2022;.
[39] Brooks BR, Brooks III CL, Mackerell Jr AD, Nilsson L, Petrella RJ, Roux B, et al. Charmm: the biomolecular simulation program. Journal of computational chemistry 2009;30(10):1545–614.
[40] Pouvreau M, Greathouse JA, Cygan RT, Kalinichev AG. Structure of hydrated kaolinite edge surfaces: Dft results and further development of the clayff classical force field with metal–o–h angle bending terms. The Journal of Physical Chemistry C 2019;123(18):11628–38.
[41] Tersoff J. Empirical interatomic potential for carbon, with applications to amorphous carbon. Physical Review Letters 1988;61(25):2879.
[42] Brenner DW. Empirical potential for hydrocarbons for use in simulating the chemical vapor deposition of diamond films. Physical review B 1990;42(15):9458.
[43] Brenner DW, Shenderova OA, Harrison JA, Stuart SJ, Ni B, Sinnott SB. A second-generation reactive empirical bond order (rebo) potential energy expression for hydrocarbons. Journal of Physics: Condensed Matter 2002;14(4):783.
[44] Van Duin AC, Dasgupta S, Lorant F, Goddard WA. Reaxff: a reactive force field for hydrocarbons. The Journal of Physical Chemistry A 2001;105(41):9396–409.
[45] Buehler MJ, Van Duin AC, Goddard III WA. Multiparadigm modeling of dynamical crack propagation in silicon using a reactive force field. Physical review letters 2006;96(9):095505.
[46] Cranford SW, Buehler MJ. Mechanical properties of graphyne. Carbon 2011;49(13):4111–21.
[47] Zhang YA, Tao J, Chen X, Liu B. Mixed-pattern cracking in silica during stress corrosion: A reactive molecular dynamics simulation. Computational materials science 2014;82:237–43.
[48] Hou D, Ma H, Zhu Y, Li Z. Calcium silicate hydrate from dry to saturated state: Structure, dynamics and mechanical properties. Acta materialia 2014;67:81–94.
[49] Rimsza JM, Jones RE, Criscenti LJ. Crack propagation in silica from reactive classical molecular dynamics simulations. Journal of the American Ceramic Society 2018;101(4):1488–99.
[50] Pitman MC, Van Duin AC. Dynamics of confined reactive water in smectite clay–zeolite composites. Journal of the American Chemical Society 2012;134(6):3042–53.
[51] Skipper NT, Chang FRC, Sposito G. Monte carlo simulation of interlayer molecular structure in swelling clay minerals. 1. methodology. Clays and Clay minerals 1995;43(3):285–93.
[52] van Duin A. Personal communications 2022;.
[53] Evans DJ, Holian BL. The nose–hoover thermostat. The Journal of chemical physics 1985;83(8):4069–74.
[54] Rappe AK, Goddard III WA. Charge equilibration for molecular dynamics simulations. The Journal of Physical Chemistry 1991;95(8):3358–63.
[55] Zhou X, Zimmerman J, Reedy Jr E, Moody N. Molecular dynamics simulation based cohesive surface representation of mixed mode fracture. Mechanics of Materials 2008;40(10):832–45.
[56] Stukowski A. Visualization and analysis of atomistic simulation data with ovito–the open visualization tool. Modelling and simulation in materials science and engineering 2009;18(1):015012.
[57] Marry V, Rotenberg B, Turq P. Structure and dynamics of water at a clay surface from molecular dynamics simulation. Physical Chemistry Chemical Physics 2008;10(32):4802–13.
[58] Siever R, Woodford N. Sorption of silica by clay minerals. Geochimica et Cosmochimica Acta 1973;37(8):1851–80.
[59] He H, Tao Q, Zhu J, Yuan P, Shen W, Yang S. Silylation of clay mineral surfaces. Applied clay science 2013;71:15–20.
[60] Lee JH, Guggenheim S. Single crystal x-ray refinement of pyrophyllite-1 tc. American Mineralogist 1981;66(3-4):350–7.
[61] Rice J, Paris P, Merkle J. Some further results of j-integral analysis and estimates. In: Progress in flaw growth and fracture toughness testing. ASTM International; 1973,.
[62] Lakshmikantha MR. Experimental and theoretical analysis of cracking in drying soils 2009;.
[63] Chandler H. The use of non-linear fracture mechanics to study the fracture properties of soils. Journal of Agricultural Engineering Research 1984;29(4):321–7.
[64] Harison JA, Hardin BO, Mahboub K. Fracture toughness of compacted cohesive soils using ring test. Journal of geotechnical engineering 1994;120(5):872–91.
[65] Spray JG. Frictional melting processes in planetary materials: From hypervelocity impact to earthquakes. Annual Review of Earth and Planetary Sciences 2010;38:221–54.
[66] Liu Y. Fracture toughness assessment of shales by nanoindentation 2015;.
[67] Anderson TL. Fracture mechanics: fundamentals and applications. CRC press; 2017.